\documentclass[twocolumn]{aastex61}

\newcommand{\Msol}{\mbox{$M_{\sun}$}}
\newcommand{\Msolxyr}{\mbox{$M_{\sun}$~yr$^{-1} / 10$~km~s$^{-1}$}}

\newcommand{\kms}{\mbox{km s$^{-1}$}}
\newcommand{\muas}{\mbox{$\mu$as}}

\newcommand{\muJb}{\mbox{$\mu$Jy~beam$^{-1}$}}

\newcommand{\rrev}{\mbox{$r_{\rm rev.\; shock}$}}
\newcommand{\thfl}{\mbox{$\theta_{\rm90\%\;flux}$}}
\newcommand{\sshell}{{\rm shell}}
\newcommand{\scomp}{{\rm comp}}
\newcommand{\sEM}{{\rm EM}}
\newcommand{\cmsixpc}{\mbox{cm$^{-6}$~pc}}
\newcommand{\LR}{\mbox{$L_{\rm R}$}}
\newcommand{\LX}{\mbox{$L_{\rm X}$}}
\newcommand{\nupeak}{\mbox{$\nu_{\rm peak}$}}
\newcommand{\nuinv}{\mbox{$\nu_{\rm inv}$}}
\newcommand{\Mion}{\mbox{$M_{\rm ion}$}}
\newcommand{\phneg}{\phantom{-}}

\defcitealias{SN86J-1}{Paper I} 
\defcitealias{SN86J-2}{Paper II}
\defcitealias{SN86J-3}{Paper III}

\shortauthors{Bietenholz \& Bartel}

\begin{document}
      

\title{SN 1986J VLBI. IV.  The Nature of the Central Component}

\author{Michael F. Bietenholz}\affiliation{Department of Physics
  and Astronomy, York University, Toronto, M3J~1P3, Ontario, Canada}
\affiliation{Hartebeesthoek Radio Observatory, PO Box 443,
  Krugersdorp, 1740, South Africa}

\author{Norbert Bartel}\affiliation{Department of Physics
  and Astronomy, York University, Toronto, M3J~1P3, Ontario, Canada}

\accepted{to the {\em Astrophysical Journal}}

\begin{abstract}
We report on VLA measurements between 1 and 45~GHz of the evolving
radio spectral energy distribution (SED) of SN~1986J, made in
conjunction with VLBI imaging.  The SED of SN~1986J is unique among
supernovae, and shows an inversion point and a high-frequency
turnover. Both are due to the central component seen in the VLBI
images, and both are progressing downward in frequency with time.  The
optically-thin spectral index of the central component is almost the
same as that of the shell.  We fit a simple model to the evolving SED
consisting of an optically-thin shell and a partly-absorbed central
component. The evolution of the SED is consistent with that of a
homologously expanding system.  Both components are fading, but the
shell more rapidly.  We conclude that the central component is
physically inside the expanding shell, and not a surface hot-spot
central only in projection.  Our observations are consistent with the
central component being due to interaction of the shock with the dense
and highly-structured circumstellar medium that resulted from a period
of common-envelope evolution of the progenitor.  However a young
pulsar-wind nebula or emission from an accreting black hole can also
not be ruled out at this point.
\end{abstract}


\section{Introduction}
\label{sintro}

SN 1986J was one of the most radio luminous supernovae ever observed,
and one of the few supernovae still detectable more than $t = 30$
years after the explosion, thus we have been able to follow its
evolution for longer than was possible for most other SNe.  Its radio
brightness and relatively close distance allowed both resolved very
long baseline interferometry (VLBI) images and accurate estimations of
its broadband radio spectral energy distribution (SED) to be made.
This paper is the fourth in our series of papers on SN~1986J,
\citet{SN86J-1, SN86J-2} and \citet{SN86J-3}, which we will refer to
as Papers I to III respectively. In this fourth paper in the series,
we discuss mainly the evolution of the broadband radio SED, but with
reference to the morphology as seen in the VLBI images.

For the convenience of the reader, we repeat some of the introductory
material from \citetalias{SN86J-3} here.  SN~1986J was first
discovered in the radio, some time after the explosion
\citep{vGorkom+1986, Rupen+1987}. The best estimate of the explosion
epoch is $1983.2 \pm 1.1$ \citep[$t = 0$, Paper I; see
  also][]{Rupen+1987, Chevalier1987, WeilerPS1990}\nocite{SN86J-1}.
It occurred in the nearby galaxy NGC~891, for whose distance the
NASA/IPAC Extragalactic Database (NED) lists 19 measurements with a
mean of $10.0 \pm 1.4$~Mpc, which value we adopt throughout this
paper.

Optical spectra, taken soon after the discovery, showed a somewhat
unusual spectrum with narrow linewidths, but the prominent H$\alpha$
lines led to a classification as a Type~IIn supernova
\citep{Rupen+1987}.
It is an unusually long-lasting SN at all wavelengths, and has been
detected in the optical \citep{Milisavljevic+2008}, infrared
\citep{Tinyanont+2016} and X-ray \citep{Houck2005a} more than two
decades after the explosion.  It has been observed with VLBI since
1987, and we refer the interested reader to a sequence of VLBI images
in \citetalias{SN86J-3} which show both the expansion and the
non-selfsimilar evolution over almost three decades.

The structure seen in the VLBI images shows an expanding, albeit
somewhat distorted shell, but also two strong compact enhancements of
the brightness: one to the NE of the shell center, and a second at or
very near the projected center.  Such a central radio component has not so
far been seen in any other supernova \citep[see
  e.g.,][]{SNVLBI_Cagliari, BartelB2014IAUS}, making SN~1986J a
particularly interesting case.

\begin{figure*}
\centering
\includegraphics[width=0.7\linewidth]{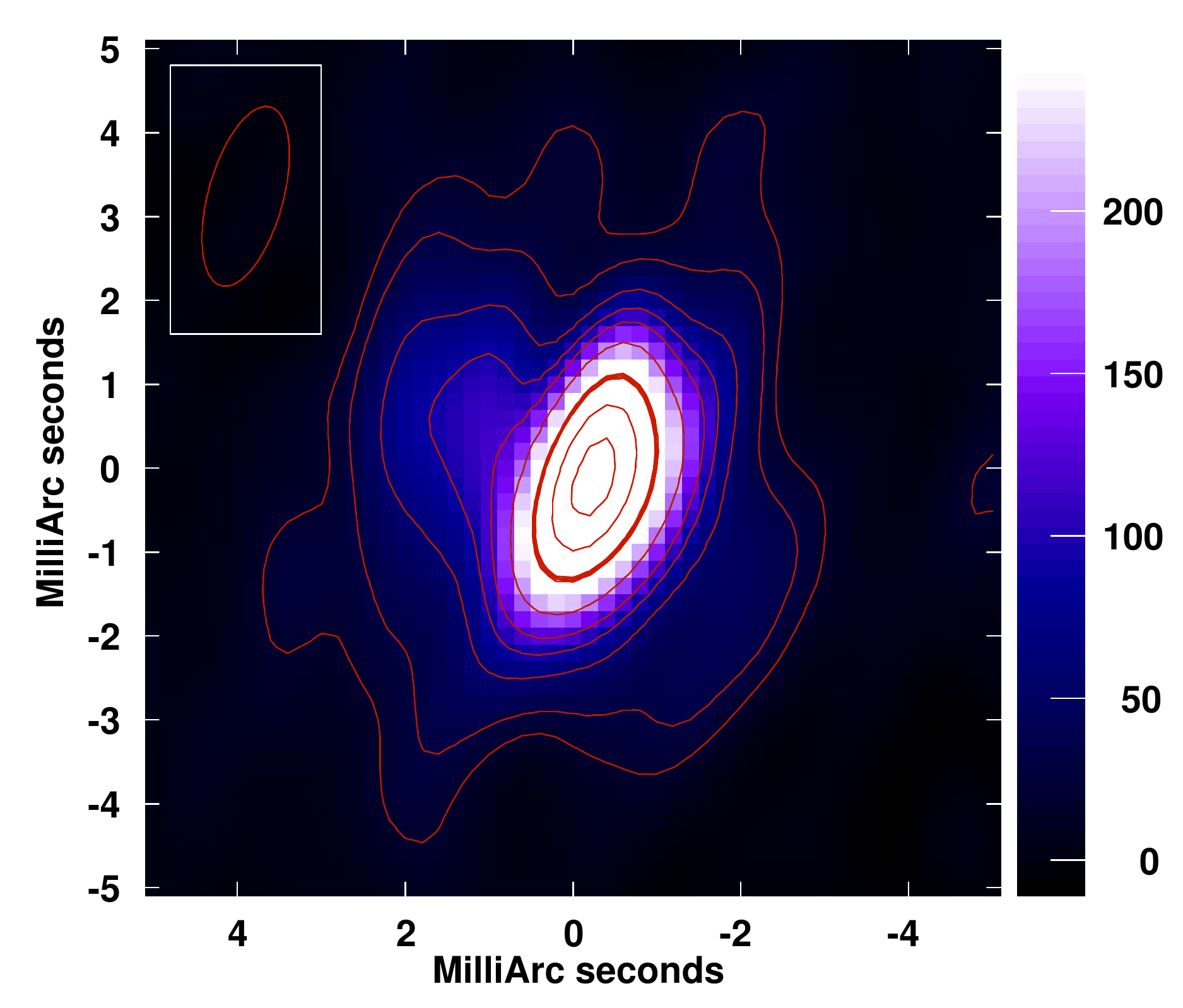}
\caption{The 5-GHz VLBI image of SN~1986J made from observations on
  2014 Oct.\ 23, at age 31.6~yr, reproduced from \citetalias{SN86J-3}.
  The contours are drawn at $-3$, 3, 5, 10, 15, 20, 30, {\bf 50}, 70
  and 90\% of the peak brightness, with the 50\% contour being
  emphasized.  The peak brightness was 617~\muJb, the total CLEANed
  flux density was 1622~$\mu$Jy, and the background rms brightness was
  5.9~\muJb.  The color-scale is labeled in \muJb, and is saturated so
  as to better show the low-level emission.  North is up and east to
  the left, and the FWHM of the convolving beam of 2.21~mas $\times \;
  0.89$~mas at p.a.\ $-15$\arcdeg\ is indicated at upper left.}
\vspace{0.1in} 
\label{fimg}
\end{figure*}

The radio emission from a supernova is synchrotron emission, and
therefore the broadband SED reflects the underlying distributions of
relativistic particles and free-free absorbing thermal matter, and
holds important clues to the physics.  The evolution of SN~1986J's SED
was quite unusual, as we discussed in \citetalias{SN86J-2}.  Usually,
the SED of a SN evolves in a fairly predictable way: the flux density,
$S_\nu$, at frequency, $\nu$, shows an optically thick rise at low
frequencies up to a turnover frequency, and an optically-thin
powerlaw above the turnover frequency, with values of the spectral
index, $\alpha$, ($S_\nu \propto \nu^\alpha$) usually in the range of
$-0.5$ to $-0.8$.  The turnover frequency evolves downward in time,
typically reaching 10~GHz at $t \sim 100$~d \citep[see,
  e.g.,][]{Weiler+2002}.  There is, however, a large variation in this
timescale: for SN~1987A, for example, the turnover frequency reached
10~GHz at $t < 1$~d \citep{Turtle+1987} while for SN~2001em
\citep[e.g.,][]{SN2001em-2}, as well as for SN~1986J
\citep{WeilerPS1990}, it did not reach 10~GHz till $t \sim 1000$~d.

Other than the relatively slow turn-on, and its high radio luminosity,
the evolution of SN~1986J's SED was unremarkable till 1998\@.  At that
time, an inversion appeared in the spectrum, with the brightness
increasing with increasing frequency above $\sim$10~GHz, up to a
high-frequency turnover at $\sim$20~GHz \citepalias{SN86J-1}.  In
\citet{SN86J-Sci}, we showed by means of phase-referenced
multi-frequency VLBI imaging, that this spectral inversion was
associated with a bright, compact component in the projected center of
the expanding shell. At that time, (in late 2002 or at $t \sim
20$~yr), the central component was clearly present in the 15~GHz
image, but not discernible in the 5~GHz one.  Since then, the central
component has become bright also at 5~GHz \citepalias{SN86J-2}, and it
dominates the 5-GHz image from 2014 \citepalias{SN86J-3}, observing
code GB074, which we show in Figure.~\ref{fimg}. The inversion in the
SED almost certainly represents radio emission which is partly
absorbed.  Although both synchrotron self-absorption (SSA) and
free-free absorption (FFA) are seen in SNe, we argue in
\citetalias{SN86J-1} that SSA is not plausible in this case, and the
absorption can therefore be ascribed to FFA from thermal material
along the line of sight.

In \citetalias{SN86J-3}, we discussed the evolution of the VLBI
images, while in this paper we discuss the evolution of the broadband
spectral energy distribution as determined mostly from measurements
with the National Radio Astronomy's\footnote{The National Radio
  Astronomy Observatory, NRAO, is a facility of the National Science
  Foundation operated under cooperative agreement by Associated
  Universities, Inc.} (NRAO) Karl G. Jansky Very Large Array (VLA) as
well as with VLBI, and discuss how the evolving SED relates to the
features seen in the VLBI images and what it can tell us about the
nature of the SN.

\pagebreak[4]
\section{VLA Observations and Data Reduction}
\label{svla}

We obtained multi-frequency VLA observations to measure SN~1986J's
total flux density at a range of frequencies at several different
epochs.  The dates and frequencies are given in Table~\ref{tvla}, and
the observing codes were 11A-130, 12B-256 for the 2011 June and 2012
April VLA runs respectively.

\begin{deluxetable}{c@{\hspace{0.2in}}c@{\hspace{0.2in}}D@{\hspace{0.2in}}C}
\tabletypesize{\footnotesize}
\tablecaption{Multi-frequency flux density measurements of SN 1986J\label{tvla}}
\tablehead{
\colhead{Date} & \colhead{Age\tablenotemark{a}} & 
\multicolumn2c{Frequency} & \colhead{Flux density\tablenotemark{b}} \\
  & \colhead{(yr)}  & \multicolumn2c{(GHz)} & \colhead{(mJy)}
}
\decimals
\startdata
2011 Jun 9  & 28.2 &  4.89 &  1.76 \pm 0.14 \\
   ``       & 28.2 &  7.80 &  2.49 \pm 0.13 \\
   ``       & 28.2 & 22.46 &  2.27 \pm 0.12 \\
   ``       & 28.2 & 33.44 &  1.99 \pm 0.21 \\
   ``       & 28.2 & 43.28 &  1.66 \pm 0.20 \\
2011 Jun 18 & 28.3 &  1.45 &  1.93 \pm 0.14 \\
   ``       & 28.3 &  1.82 &  1.66 \pm 0.12 \\
   ``       & 28.3 &  3.15 &  1.49 \pm 0.09 \\
   ``       & 28.3 &  4.89 &  1.79 \pm 0.10 \\
2012 Apr 10 & 29.6 &  1.10 &  1.61 \pm 0.10 \\
   ``       & 29.6 &  1.40 &  1.34 \pm 0.08 \\
   ``       & 29.6 &  1.65 &  1.37 \pm 0.08 \\
   ``       & 29.6 &  1.87 &  1.21 \pm 0.07 \\
   ``       & 29.6 &  2.38 &  1.23 \pm 0.07 \\
   ``       & 29.6 &  3.03 &  1.30 \pm 0.07 \\
   ``       & 29.6 &  3.69 &  1.43 \pm 0.07 \\
   ``       & 29.6 &  4.99 &  1.82 \pm 0.09 \\
   ``       & 29.6 &  5.96 &  2.11 \pm 0.11 \\
   ``       & 29.6 &  8.74 &  2.56 \pm 0.13 \\
   ``       & 29.6 &  9.56 &  2.63 \pm 0.13 \\
   ``       & 29.6 & 13.37 &  2.98 \pm 0.15 \\
   ``       & 29.6 & 14.63 &  2.97 \pm 0.15 \\
   ``       & 29.6 & 20.70 &  2.50 \pm 0.17 \\
   ``       & 29.6 & 21.70 &  2.27 \pm 0.19 \\
   ``       & 29.6 & 32.00 &  2.02 \pm 0.11 \\
   ``       & 29.6 & 41.00 &  1.37 \pm 0.14 \\
2014 Oct 22 & 31.6 &  5.00 &  1.62 \pm 0.16\tablenotemark{c} \\
\enddata
\tablenotetext{a}{The age of SN 1986J, taken with respect to an explosion
epoch of 1983.2 \citepalias[see][]{SN86J-1}.}
\tablenotetext{b}{The flux density and its 1$\sigma$ standard error,
  with the latter including both the statistical and systematic
  contributions.  All flux densities are from VLA observations unless
  noted.}
\tablenotetext{c}{Flux density determined from VLBI, as described in
  \citetalias{SN86J-3}, rather than from VLA observations.\\}
\end{deluxetable}

The VLA data were reduced following standard procedures using AIPS for
the 2011 data and both AIPS and CASA for the 2012 data.  The
flux-density scales were calibrated by using observations of the
standard flux-density calibrators 3C~286 and 3C~48 on the scale of
\citet{Baars+1977} and \citet{PerleyB2013a}.  An atmospheric opacity
correction using mean zenith opacities was applied, and
NGC~891/SN~1986J was self-calibrated in phase to the extent permitted
by the signal-to-noise ratio for each epoch and frequency.  All our
VLA measurements had resolutions $<2\arcsec$, allowing a reliable
separation of SN~1986J's emission from the extended emission from the
galaxy.  We measured flux densities by fitting an elliptical Gaussian
to the image, with a zero-level also being fit in cases where there
was significant background emission from the galaxy.\footnote{We note
  that \citet{Mulcahy2014} gives a flux density of 8.8 mJy for
  SN~1986J at 146~MHz, measured with LOFAR in late 2012, which is
  consistent with, although slightly higher than the extrapolation of
  our spectrum at $t = 29.6$~yr.  Separation of the SN emission from
  that of the galaxy might have been difficult at their resolution of
  25\arcsec.}  measured flux densities in Table~\ref{tvla}.

\begin{figure}
\centering
\includegraphics[width=\linewidth]{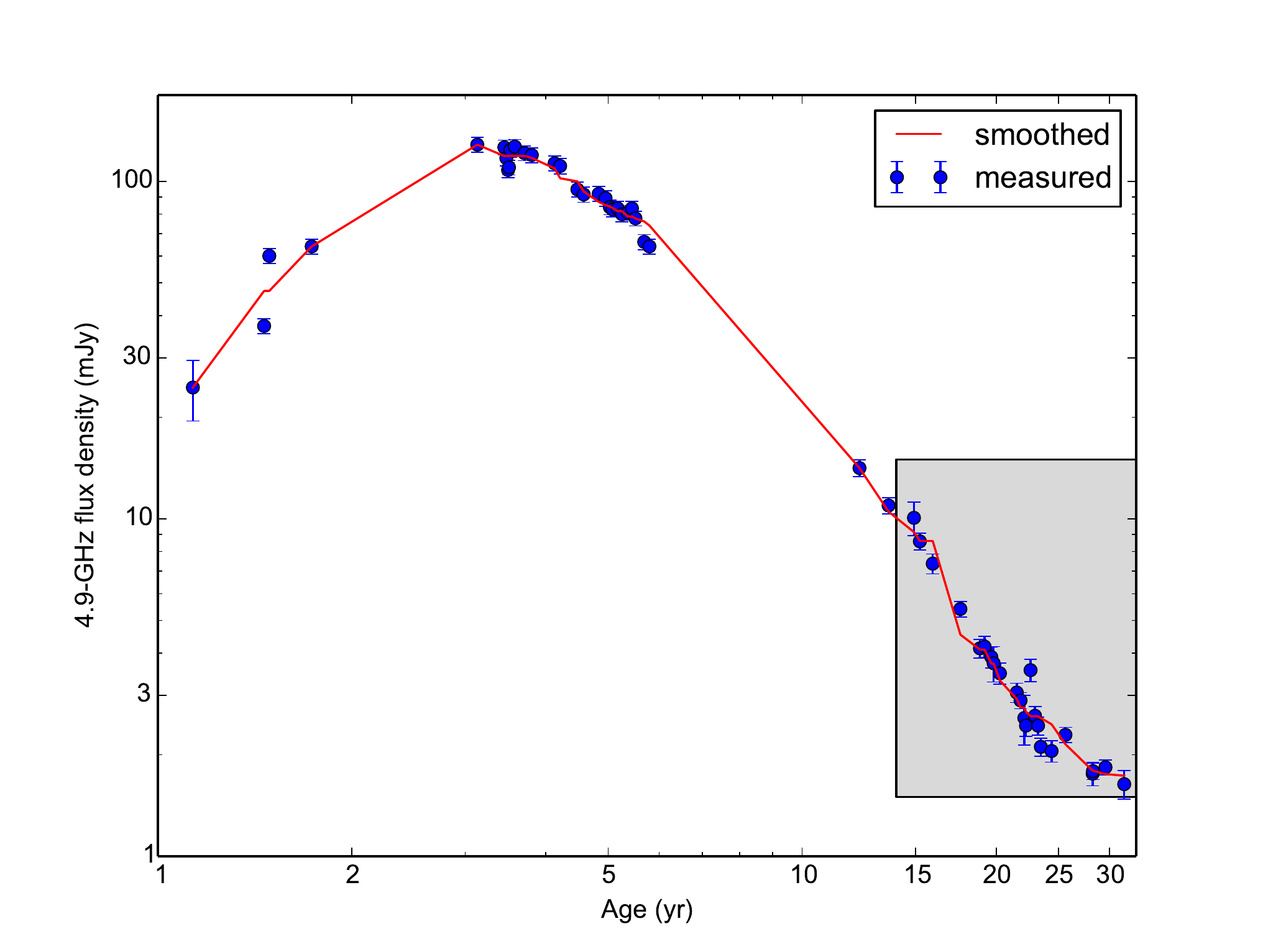}
\caption{The complete 5-GHz lightcurve of SN~1986J, based on
  measurements from the present work as well as ones from
  \citet{WeilerPS1990}, \citetalias{SN86J-1} and \citetalias{SN86J-2}.
  We show the measured values with their estimated standard errors in
  blue, and a smoothed average curve in red.  The smoothing was a
  simple boxcar averaging in log(age) and log(flux-density), with the
  width of the boxcar being 20\% in age.  The lightly shaded rectangle
  shows the time interval which we use for modeling the effect of the
  central component on the SED.
\vspace{0.3in}}  
\label{flc5}
\end{figure}

We plot first the complete 5-GHz lightcurve of SN~1986J in
Figure~\ref{flc5}, including the measurements from the present paper
as well as earlier ones \citep[from][Paper I and Paper
  II]{WeilerPS1990}\nocite{SN86J-1, SN86J-2}.  In this paper we focus
on the evolution after the emergence of the central component, that is
$t = 14$~yr, which interval is indicated in Figure~\ref{flc5} by a
lightly shaded box.  The lightcurve shows an optically thick rise till
$t \simeq 3$~yr, and an optically-thin decline thereafter.  The
decline steepens between $t = 6$ and 15~yr,
as we noted in \citetalias{SN86J-1}, the decay flattens again at $t >
25$~yr, due to the influence of the central component. Some short-term
variations are seen both at early and late times.  The peak flux
density of 128 mJy corresponds to a spectral luminosity of $1.54
\times 10^{28}$~erg~s$^{-1}$~Hz$^{-1}$.

\begin{figure*}
\centering
\includegraphics[width=0.8\linewidth]{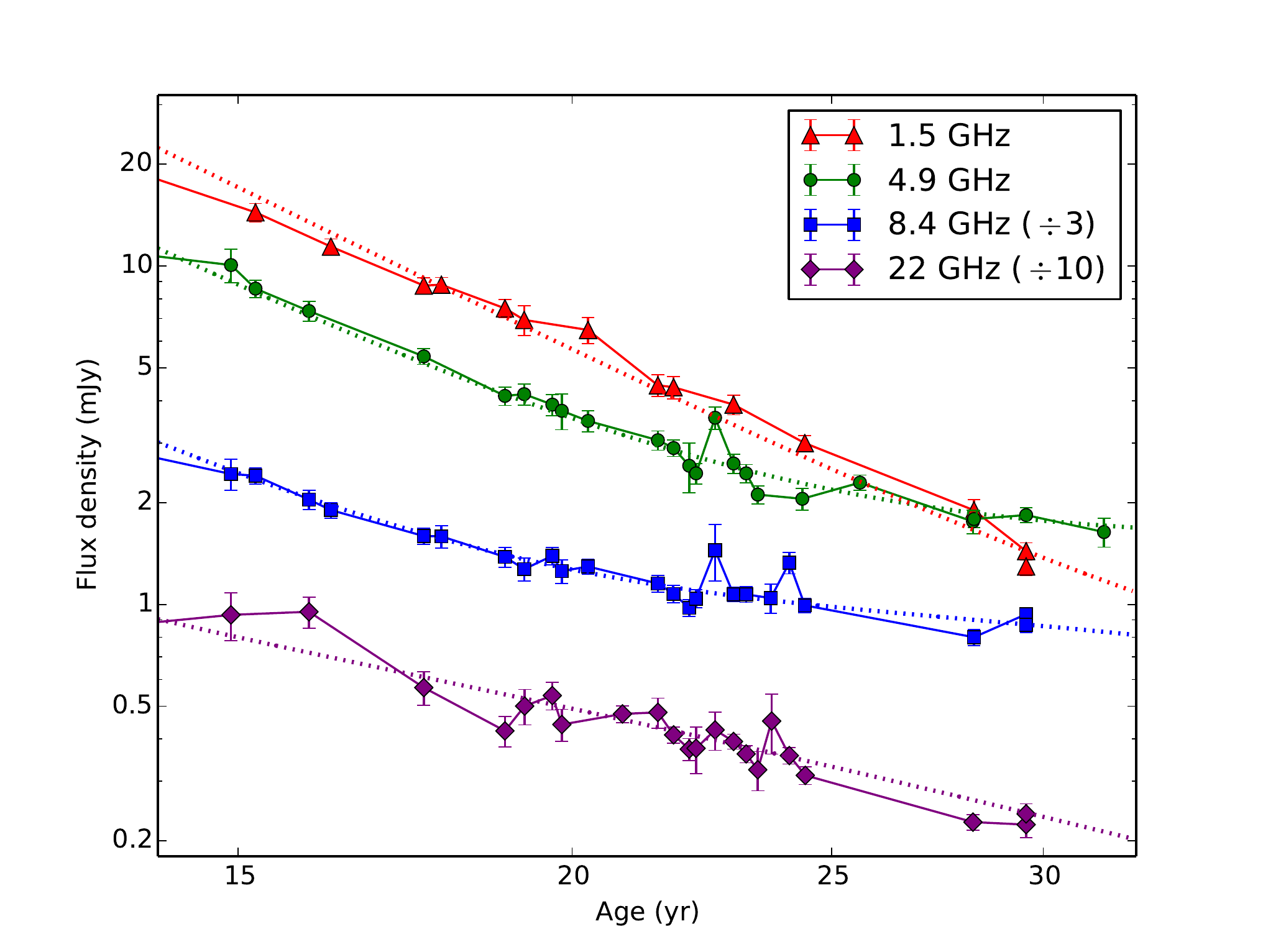}
\caption{The radio lightcurves of SN~1986J at frequencies between 1.5
  and 22~GHz. For clarity, the lightcurves at 8.4~GHz and 22~GHz were
  shifted down in flux density by factors of 3 and 10 respectively.
  The plotted error bars are $1\sigma$ standard errors which include
  both statistical and systematic contributions added in quadrature.
  Data are taken from this paper and \citetalias{SN86J-2},
  \citet{SN86J-COSPAR} and \citet{SN86J-Sci}.  In \S~\ref{sBayes}, we
  describe our Bayesian fit to the evolving spectral energy
  distribution after the first appearance of the central (inverted
  spectrum) component at $t = 14.9$~yr, and we show the fitted
  lightcurves using dotted lines of the same colors as were used for
  the measurements from the same epoch.
  \vspace{0.1in}}  
\label{flightcurve}
\end{figure*}

\begin{figure*} 
\centering
\includegraphics[width=0.8\linewidth]{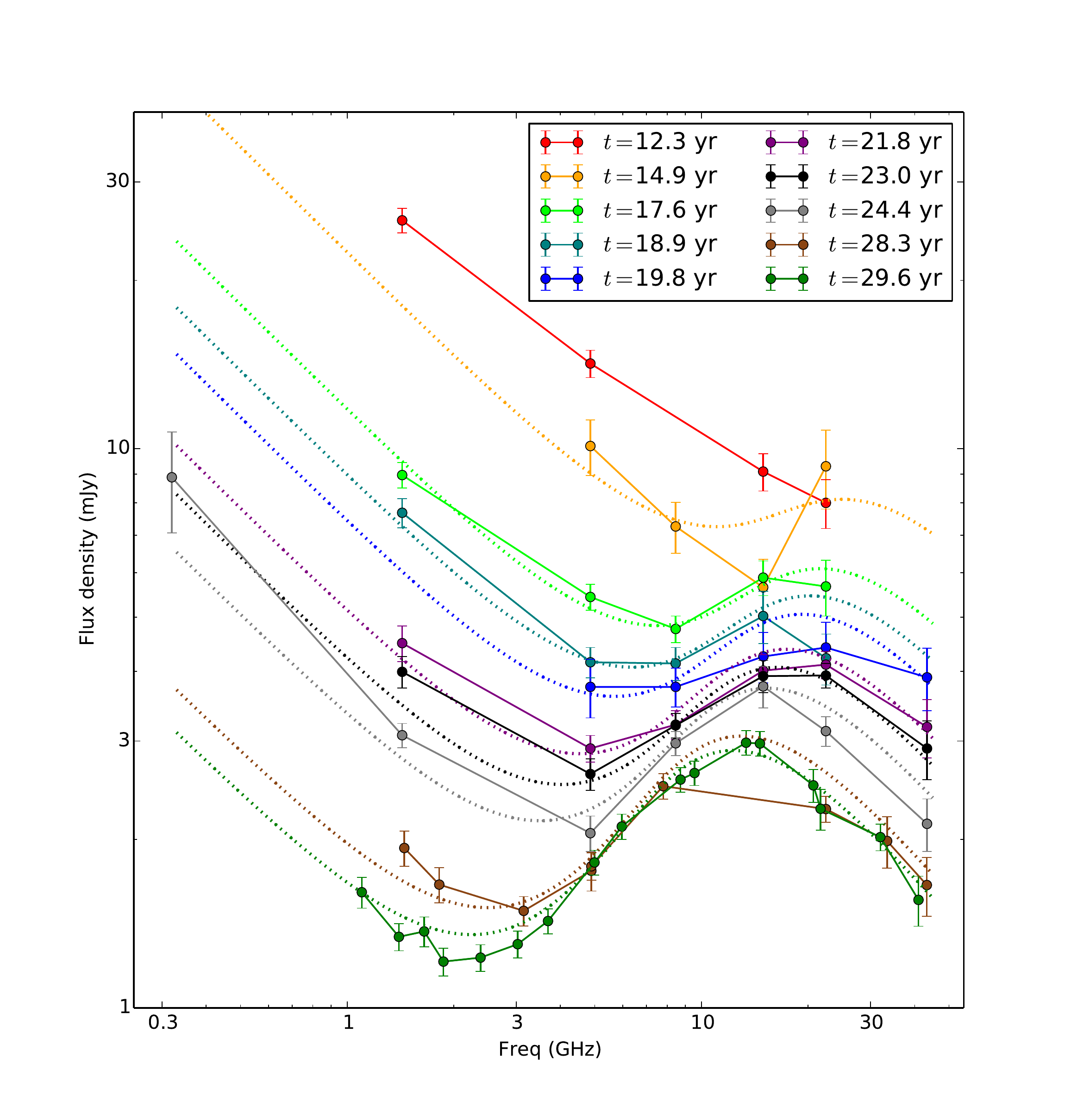}
\caption{The evolving SED of SN~1986J\@.  The colored points and
  associated solid lines show the SED at the indicated age (in years,
  calculated from $t_0 = 1983.2$), with the earliest one at the top.
  The uncertainties are estimated standard errors, with statistical
  and systematic contributions added in quadrature.  Data are taken
  from this paper, \citetalias{SN86J-2} and \citet{SN86J-Sci,
    SN86J-COSPAR}.  In \S~\ref{sBayes}, we describe our Bayesian fit
  to the evolving spectral energy distribution after the first
  appearance of the central (inverted spectrum) component at $t =
  14.9$~yr.  For reference, we include the spectrum at $t=12.3$~yr,
  before the emergence of the central component, which was not used in
  our fit. We plot the fitted spectra for each subsequent epoch using
  dotted lines of the same colors as were used for the measurements
  from the same epoch.  The model is fairly simple, and fits the SEDs
  at all epochs simultaneously with only eight free parameters, so the
  fit to the measured spectrum at any one epoch is only approximate,
  but the model should give a good overall description of the
  evolution.
  \vspace{0.1in}}  
\label{fsed}
\end{figure*}

We plot the multi-frequency flux densities at $t > 14$~yr in two ways:
first, in Figure~\ref{flightcurve} we plot the evolution of the flux
density with time at several different frequencies for the period 1998
to 2016 ($t = 15$ to 32~yr).  Second, in Figure~\ref{fsed} we plot the
SED of a number of recent epochs.  In both figures, we include some
earlier measurements from \citetalias{SN86J-2}, and \citet{SN86J-Sci,
  SN86J-COSPAR}.

\section{Bayesian Fit to the Evolution of the Spectral Energy Distribution}
\label{sBayes}

\subsection{Model for the Bayesian Fit}
\label{smodel}

Our latest broadband SED shows that at $t = 29.6$~yr, the
high-frequency turnover, \nupeak, occurs at $\sim$15~GHz, while the
inversion in the spectrum, \nuinv, occurs at $\sim$2~GHz.  The SED
from $t = 29.6$~yr as well as the earlier ones are shown in
Figure~\ref{fsed}.  The high-frequency turnover was not observed till
$t = 17.6$~yr, at which point $\nupeak \simeq 17$~GHz.  By $t =
29.6$~yr, \nupeak\ had moved downwards but only slightly to
$\sim$14~GHz, and so $\nupeak\ \propto t^{(-0.3\pm0.3)}$.
The inversion was first detected at $t = 14.9$~yr at which time
$\nuinv \simeq 15$~GHz.  By $t = 29.6$~yr, $\nuinv \simeq 2$~GHz, so
the inversion point is moving downwards in frequency much more rapidly
with $\nuinv \propto t^{(-2.9 \pm 0.4)}$.

We fitted a simple model to the observations after $t = 14$~yr, where
the SED is modeled as a combination of two parts.  The first, due to
the shell, is assumed to suffer no absorption so:
$$S_\sshell = S_{0,\sshell} \left(\frac{t}{\rm 20 \> yr}\right)^{b_\sshell} \,
\left(\frac{\nu}{\rm 1 \> GHz}\right)^{\alpha_\sshell}$$ 
The second, due to the central component, has an intrinsic SED of
$$S_\scomp = S_{0,\scomp} \left(\frac{t}{\rm 20 \>
  yr}\right)^{b_\scomp} \> \left(\frac{\nu}{\rm 1 \>
  GHz}\right)^{\alpha_\scomp}$$ but is free-free absorbed by material
with an emission measure, EM $= \int {N_e^2 \cdot dl}$, where $N_e$ is
the electron number density, and $l$ is the path-length along the line
of sight.

In our model, the EM is also time-variable and is given by
$$ \mathrm{EM} = \mathrm{EM}_0 \cdot \left(\frac{t}{\rm 20 \ yr}\right)^{b_\sEM} \, \cmsixpc,$$
and the corresponding optical depth is
$$\tau    =  3.28 \times 10^{-7} \, 
  \left(\frac{\nu}{\mathrm{GHz}}\right)^{-2.1} \, 
  \left(\frac{T_e}{10^4 \, \mathrm{K}}\right)^{-1.35}\, 
  \left(\frac{\mathrm{EM}}{\cmsixpc}\right).$$
We assume an electron temperature of $T_e = 10^4$~K\@. 

In the case of purely external absorption, the inverted part of the
spectrum has a steep frequency dependence with
$S_{\mathrm{comp,\,abs}} \propto \nu^{2.1}$, which does not well match
our observed SEDs \citepalias[see Figure~\ref{fsed} and][]{SN86J-2}.
We therefore take the emitting and the absorbing material to be mixed,
in which case the inverted part of the spectrum has $S_{\mathrm{comp,\,
    abs}} \propto \nu^{(\alpha_\scomp + 2.1)}$.
The contribution of the partly absorbed central component is then
$$S_{\mathrm{comp,\,abs}} = S_\scomp \, \tau^{-1} \, (1 - e^{-\tau})$$

This model is almost certainly an oversimplification, for example the
spectral indices might also be time-variable, or the time-dependence
of the various quantities might not in fact be of a powerlaw form.
However, the results give some physical insight into how the
time-dependencies of various quantities might interact to produce the
observed sequence of SEDs.

\pagebreak[4]
\subsection{Results from the Bayesian Fit}
\label{sBresults}

Having thus defined our model, we then performed a Bayesian fit, and
used Monte-Carlo Markov chain integration to estimate the posterior
probability distribution of the eight free parameters, namely
$S_{0,\sshell}$, $b_\sshell$, $\alpha_\sshell$, $S_{0,\,\scomp}$,
$b_\scomp$, $\alpha_\scomp$, EM$_0$, and $b_\sEM$.  Further details on
the Bayesian fit and a comparison to a least-squares fit are given in
Appendix~\ref{aBaylsq}.

We obtained the following values:\\
\begin{tabular}{r r l}
~ & $S_{0,\sshell} =   $&$ \phn 7.07 \pm 0.17$ mJy, \\[-3pt]
  & $b_\sshell =      $&$ -3.92 \pm 0.07$, \\[-3pt]
  & $\alpha_\sshell = $&$ -0.63 \pm 0.03$, \\[-3pt]
  & $S_{0,\scomp} =    $&$ 61 \pm 17$~mJy,\\[-3pt]
  & $b_\scomp =       $&$ -2.07 \pm 0.16$, \\[-3pt]
  & $\alpha_\scomp =  $&$ -0.76 \pm 0.07$, \\[-3pt]
  & EM$_0 =          $&$ (1.64 \pm 0.20) \times 10^9$~\cmsixpc, and \\[-3pt]
  & $b_\sEM =        $&$ -2.72 \pm 0.26$.  
\end{tabular}\\
The~listed uncertainties are the standard deviations over the
posterior distribution as determined by the Markov-Chain Monte-Carlo
integration.  We plot the fitted spectra (as the dotted lines) in
Figure~\ref{fsed}.  Note that $S_{0,\scomp}$ and $\alpha_\scomp$ are
highly anti-correlated (see Appendix~\ref{aBaylsq}).  The fitted flux
density of the central component at 10~GHz is better constrained than
$S_{0,\scomp}$, and is $10.4 \pm 1.1$~mJy as evaluated from the
posterior distribution.  Our fit suggests that the shell brightness is
declining rapidly, with $S_\sshell \propto t^{-3.92\pm0.07}$.  Even
the central component is declining in unabsorbed flux density, with
$S_\scomp \propto t^{-2.07\pm0.16}$.  As can be seen in
Fig.~\ref{flightcurve}, the 22-GHz flux density, where the central
component is mostly optically thin, continues to decline, albeit
slightly more slowly than that at 1.5~GHz, which is likely still
dominated by the shell.  At 10~GHz and $t = 20$~yr, the unabsorbed
flux density of the central component is $6 \pm 1$ times larger than
that of the shell.  The central component is declining significantly
less rapidly than the shell which implies that it is becoming
relatively more dominant.  In other words the fraction of the emission
due to the central component is increasing with time, even at
frequencies where the central component is optically thin.  Once the
central component dominates, the total flux density decline would
flatten to about $S \propto t^{-2.07}$.

The amount of absorption suffered by the central component is also
decreasing with time: in our fitted model, the emission measure, EM,
of the material responsible for the free-free absorption of the
central component is EM~$\propto t^{-2.72\pm0.26}$.  It is the
declining absorption that produces the apparent increase with time of
the central component's flux density at some frequencies, in
particular at 5~GHz where it has come to dominate the images in the
last decade.

Since EM = $\int {N_e^2 \cdot dl}$, in a system which has a constant
total number of free electrons and is homologously expanding $\propto
t^q$ (all spatial dimensions $\propto t^q$), one would expect EM
$\propto t^{-5q}$.  Our fitted value of $b_\sEM$ therefore suggests a
system expanding with spatial dimensions $\propto t^{0.54\pm0.05}$.
This indicates a system which is slightly more decelerated than the
supernova shell, for which we measured an expansion $r_{\rm out}
\propto t^{0.69 \pm 0.03}$ \citepalias{SN86J-2}.\footnote{We
  determined the expansion from measurements up to $t = 25.6$~yr in
  \citetalias{SN86J-2}.  In the latest VLBI observations at
  $t=31.6$~yr, as discussed in \citetalias{SN86J-3}, we could no
  longer reliably determine the shell size because of the low
  signal-to-noise and changing brightness distribution.  A continued
  expansion with $r \propto t^{0.69}$ is compatible with the
  measurements, although increased deceleration, for example $r
  \propto t^{0.54}$, is also possible.  For our purposes here, we
  assume a continued steady deceleration ($\propto t^{0.69}$) through
  to $t = 31.6$~yr.}

From our VLBI images, we obtained a FWHM diameter of
$1.3\times10^{17}$~cm or 0.044~pc for the central component at
$t=31.6$~yr.  Our fit to the SED suggests that at that time, EM $=4.7
\times 10^{8}$~\cmsixpc, implying an average $N_e =
1.0\times10^{5}$~cm$^{-3}$ if the absorption is internal to the
central component.  For a uniform sphere of fully ionized material
with mass in u (unified atomic mass units) per free
electron\footnote{$\mu_e$ depends on the composition of the ionized
  material, being approximately $2/(1+X)$ where $X$ is the fraction of
  hydrogen by mass.  For fully ionized material of solar composition,
  such as the outer envelope of the star, $\mu_e = 1.15$. For fully
  ionized processed material, such as the interior layers of ejecta,
  $\mu_e \simeq 2$.  The progenitor of SN~1986J is expected to have
  retained a large portion of its H-rich envelope at the time of
  explosion, and we therefore adopt the intermediate value of $\mu_e =
  1.3$, somewhat higher than for pure envelope material but not as
  high as for the interior of the star.}, of $\mu_e = 1.3$, this
amounts to $M_{\rm ionized} \gtrsim 0.14$~\Msol, so the mass within
the central component would be at least that amount (see \S \ref{sabs}
below for a discussion of the case where the absorption is not
internal to the central component but rather due to the supernova
ejecta).

The fitted optically-thin spectral indices for the shell and the
central component were $\alpha_\sshell = -0.63 \pm 0.03$ and
$\alpha_\scomp = -0.76 \pm 0.07$.  Although formally these are
different by $1.7\sigma$, considering the relatively simple nature of
our model, it is not clear whether the difference is significant. We
performed a fit to the evolving SED where we used a single spectral
index, $\alpha$, for both the shell and the central component, which
fit the measurements almost as well, and obtained $\alpha = -0.64 \pm
0.03$.  Our measurements are therefore compatible with both the shell
and the central component having the same optically-thin spectral
index.

Our Bayesian fit to the SEDs implies that $S_\scomp = 14 \times
S_\sshell$ at $t=29.6$~y and 5~GHz (before any absorption of the
central component).  This can be seen in Figure~\ref{fsed}, where
extrapolation of the high-frequency part of the spectrum, i.e., the
sum of shell and central component spectra, would be $\sim$14 times
higher than the low-frequency part which is just from the shell.

The approximate nature of our model notwithstanding, we can therefore
say that the unabsorbed flux density from the central component is at
least $10\times$ higher than that of the shell at $t=29.6$~yr.  From
our fitted FWHM width of the central component of $\theta_{\rm comp} =
900$~\muas\ and the estimated angular radius of the shell of $\thfl
\sim$4~mas at $t=29.6$~yr \citepalias{SN86J-3}, we calculate that the
component occupies $\sim$1.3\% of the projected area.  The central
component, therefore, must have an unabsorbed brightness $> 10/0.013$
or $>750$ times that of the shell at $t=29.6$~yr.

\subsection{Short-term Variation in the Flux Density}
\label{svariability}

The model we used for the Bayesian fit had a smoothly evolving SED.
However, as can be seen from Figures~\ref{flightcurve} and \ref{fsed},
although our model adequately reproduces the overall shapes of the
multi-frequency radio lightcurves and the evolution of the SED, the
measurements do show some significant deviations from the model.  In
particular, the lightcurve at 22~GHz, which is dominated by the
central component, seems to show larger short-term deviations from the
smoothly-evolving model, suggesting perhaps some variability of the
central component.

There is also some evidence for short-term deviations or flares at
other frequencies.  The most prominent of these is a positive
excursion of the flux density by a factor of $\sim$1.4 seen at both
4.9 and 8.4 GHz at $t = 22.6$~yr.  There was no corresponding
measurement at 1.5~GHz, and at 22-GHz only an insignificant
enhancement is seen.  
This suggests flaring behavior with a spectrum deviating notably
from a powerlaw.

\section{Discussion}
\label{sdiscuss}

We have presented new, broadband measurements of SN~1986J's SED
between 1 and 44~GHz, and fitted a model to its evolution.  How can we
interpret the evolution of the SED, and what light does it shed on the
nature of SN~1986J's mysterious central component, which appeared in
VLBI images in conjunction with a dramatic evolution in the SED?

We will proceed to discuss four different hypotheses for the origin of
the central component in light of our measurements. In
\S~\ref{sCSMclump}, we show that the first hypothesis, which is that
the central component is merely a dense condensation in the
circumstellar medium (CSM), lying by chance near the projected center
of the source, is unlikely.  Then, in \S~\ref{sabs} we discuss
absorption by the ejecta, which is a common feature of the three
remaining hypotheses for the origin of the central component. In
\S~\ref{sbinarySN} we discuss the second hypothesis, suggested by
\citet{Chevalier2012b}, in which SN~1986J was the second SN resulting
from a massive binary, where the compact remnant from the first SN
inspirals into the second star and produces a very anisotropic CSM in
which the second supernova, SN~1986J, then explodes.  In
\S~\ref{spwn}, we discuss the third hypothesis, which is that the
central component is a pulsar wind nebula around the nascent neutron
star which was left behind by the SN explosion.  Finally, in
\S~\ref{sblackhole} we discuss the fourth hypothesis, which is that
the central component is radio emission associated with a black hole
left behind by the explosion.

\pagebreak[4]
\subsection{Is the Central Component Due to a Dense Condensation in the CSM?}
\label{sCSMclump}

Is it possible that the central component is merely the result of the
supernova shock interacting with a dense condensation in the CSM,
which lies by chance close to the projected center, but on the near
side of the source along the line of sight?  If the condensation were
sufficiently dense, the condensation itself could provide the
necessary opacity to account for the inverted part of the spectrum and
the delayed turn-on. The central location would therefore be
coincidental.  Given that a somewhat similar bright radio spot was
seen to the SE in earlier images \citepalias{SN86J-1, SN86J-2}, which
is indeed thought to be due to a condensation in the CSM, such a dense
condensation would not be unique.

We measured a low proper motion for the central component, consistent
with being stationary, in \citetalias{SN86J-3}.  Such low proper
motion is expected for a CSM condensation unless its structure and
placement deviates strongly from symmetry along the line of sight.

In this scenario, one would expect the central component to fade after
the shock has progressed through it.  The central component was first
seen at $t = 14.9$~yr, when the supernova was only 60\% of its present
size, and it is still getting brighter.  Given that the condensation
would have to be quite dense, the portion of the shock traversing the
condensation would slow considerably compared to the remainder, and
for sufficiently high densities, lifetimes of a few decades would not
be excluded.  Since the central component is still brightening
relative to the shell, the shock would probably not have traversed
more than half of the putative CSM clump.  We measured a sky-plane
radius for the central component of $6.7\times10^{17}$~cm
\citepalias{SN86J-3}.  Assuming a spherical clump, and given that the
central component has been visible since $t = 14$~yr, we can calculate
that the shock speed within the clump would have to be $<1300$~\kms,
which is $\lesssim$23\% of the speed of the parts of the shock
unaffected by the central component of $\sim$5700~\kms\ \citepalias[$t
  = 15.9 - 25.6$~yr,][]{SN86J-2}.  This would necessitate quite high
densities in the clump.

The high surface brightness (before absorption) of $\gtrsim 750$ times
that of the remainder of the shell, seems hard to accommodate in this
scenario.  When the shock traverses a density jump in the CSM,
brightness changes on the same order as the magnitude of the jump in
density are expected, suggesting that the clump should have a density
$\gtrsim 750$ times that of the remainder of the CSM at that
radius.\footnote{Note that in \citetalias{SN86J-2} we estimated the
  surface brightness of the central component to be 25 times that of
  the remainder of the shell at 5 GHz.  Our present value is much
  higher because we are now considering the {\em unabsorbed}\/
  brightness of the central component.  At 5~GHz there is still
  substantial absorption, so the unabsorbed 5-GHz brightness of the
  central component is considerably higher. Furthermore, the central
  component has gotten brighter relative to the shell over the last
  few years.  These factors combined account for the large increase in
  the ratio of the brightnesses.}  Given the average densities
expected from mass loss at a rate of $4 \sim 10 \times 10^{-5}
\Msolxyr$ of $(5 \sim 13) \times 10^{22}$~g~cm$^{-3}$
\citepalias{SN86J-2}, we would therefore expect number densities in
the clump of $(3 \sim 7) \times 10^{-19}$~g~cm$^{-3}$, or, assuming
atomic H, number densities of $2 \sim 4 \times 10^{5}$~cm$^{-3}$.

A clumpy CSM has been suggested for SN~1986J with shocks driven into
dense clumps in the CSM explaining both the X-ray emission and low
line velocities \citep{Chugai1993}.  Clumps with similarly high number
densities have been suggested in the CSM of other SNe, for example
values in excess of $10^6$~cm$^{-3}$, even higher than we surmised for
a putative CSM condensation in SN~1986J, have been suggested in
SN~2006jd and SN2010jl \citep[respectively]{Smith+2009,Fransson+2014}.
However, the number of such clumps is thought to be large
($\sim$1000), their sizes small ($\sim 3\times10^{15}$~cm), and their
filling factor low, ($\sim$0.005), very much in contrast to what we
find for SN~1986J.  A clumpy CSM therefore would be unable to explain
either the fact that we see only a single bright spot in SN~1986J or
its longevity.

\subsection{Absorption by the Supernova Ejecta}
\label{sabs}

If the central component is not a fortuitously-placed clump in the
CSM, near the center only in projection, then it is in (or close to)
the physical center of the expanding supernova.  The absorption of its
radio emission at low frequencies would then be due to free-free
absorption by the intervening supernova ejecta along the line of
sight.

Can we deduce anything about the ejecta from our radio observations?
In our Bayesian model of the evolving SED we found that the best fit
had a time-dependent emission measure of 
$$ {\rm EM}(t) = (1.64 \pm 0.21) \times 10^9 \cdot \left(\frac{t}{\rm \,
  20 \, yr}\right)^{-2.72 \pm 0.26} \; \cmsixpc .$$

\subsubsection{Uniformly Distributed Free Electrons}

If we assume that the thermal free electrons responsible for the EM
are uniformly distributed out to the reverse shock, which we take to
be at 80\% of the outer radius of the SN \citepalias[taken from our
  fit in][]{SN86J-2}, and therefore to be at $4.3\times10^{17}$~cm at
$t=20$~yr, then we can calculate that the average value for $N_e$ must
be $1.1\times10^5$~cm$^{-3}$.
If we assume a uniform distribution of matter and $\mu_e = 1.3$, this
implies a mass of $\sim$40~\Msol\ of fully ionized matter.  This is
higher than is reasonable, so we must conclude that the density
distribution of the absorbing matter is non-uniform.

As shown by \citet{ChevalierF1994}, the ejecta are not expected to be
uniformly ionized, instead, probably only the
outer portion of the ejecta, nearest the reverse shock, will be highly
ionized.  The total amount of ionized material in the ejecta of a SN,
\Mion, is not well known: \Mion\ depends on the poorly-known
ionization fraction and distribution of the ionized material in the
ejecta.  It is expected that the ejecta will cool and become neutral
except for a shell which is heated by emission from the shocks
\citep{ChevalierF1994}, so \Mion\ is likely to be only a fraction of
the total ejecta mass.  The expected value of \Mion\ for SN~1986J is
likely in the range of 0.5 to 5~\Msol.  \citet{Zanardo+2014} argues
that \Mion\ in the ejecta of SN~1987A is in the range of 0.7 to
2.5~\Msol, and since SN~1986J's progenitor was probably more massive
than that of SN~1987A, we consider values of \Mion\ up to 5~\Msol, and
consider values larger than that unlikely.

\subsubsection{Non-Uniformly Distributed Free Electrons}

By making some assumptions about the shape of the ionized region
within the ejecta, we can constrain the distribution of ionized matter
using our fitted values of EM\@.  If we assume spherical symmetry,
then the ionized matter must be distributed in a region in the shape
either of a spherical shell or a sphere.  We consider both
possibilities.  First, a spherical shell, bounded on the outside at
radius \rrev\ by the reverse shock, based on the expectation that the
ejecta will be ionized from the outside by the shock.  Second, a
spherical region in the interior of the ejecta, extending out to some
radius $<\rrev$.  The EM then depends on \Mion\ and the path-length
through the ionized matter (and also, but only weakly, on $\mu_e$, and
we assume again that $\mu_e$ = 1.3 for the ejecta). The path-length is
the thickness of the shell or the radius of the sphere in the two
cases.

\begin{figure}
\centering
\includegraphics[width=\linewidth]{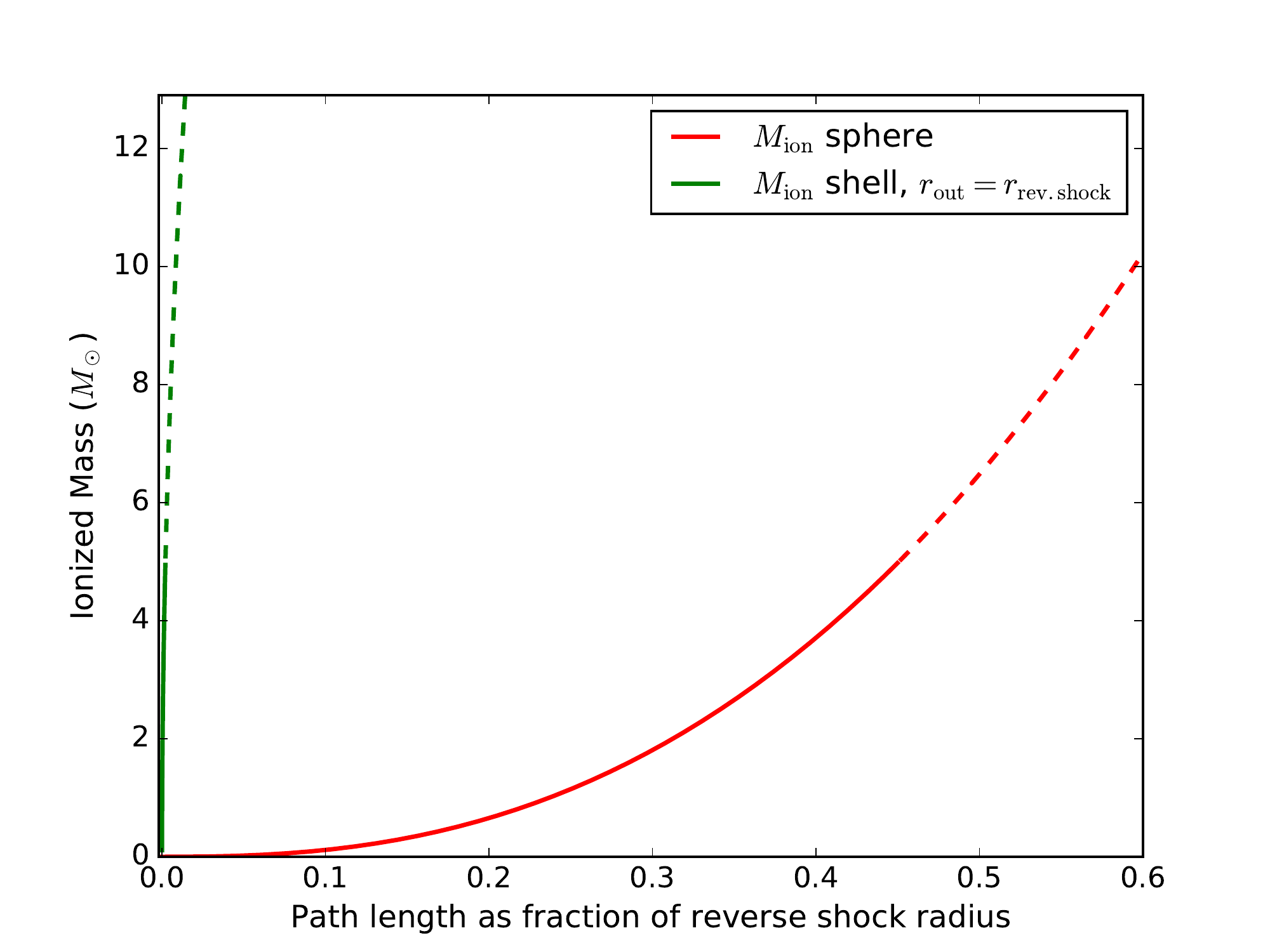}
\caption{The ionized mass (\Mion) in the ejecta of SN~1986J for an
  emission measure (EM) through the ejecta of $1.64 \times
  10^9$~\cmsixpc, as a function of path-length, which we express in
  fractions of the reverse shock radius, \rrev.  The value of EM is
  the one we found for SN~1986J at $t = 20$~yr (see text,
  \S~\ref{sBresults}).  We assume a mass per free electron, $\mu_e$ of
  1.3 u\@.  The green line near the left edge shows \Mion\ for ionized
  matter distributed in a spherical shell extending out to \rrev, with
  the path-length being the thickness of the shell, while the red line
  shows \Mion\ for a spherical distribution of ionized matter,
  extending out to so some $r < \rrev$, with the path-length being the
  radius of the sphere.  We consider \Mion\ = 5~\Msol\ to be a
  reasonable upper limit for the total amount of ionized matter in
  SN~1986J, and we plot values of \Mion $> 5$ using dashed lines.  For
  a spherical shell, only very thin shells (thickness $< 0.002 \,
  \rrev$) are possible.}
\vspace{0.1in}  
\label{fMion}
\end{figure}

In Figure~\ref{fMion} we plot the value of \Mion\ required to produce
the EM of $1.64\pm0.21 \times 10^9$~\cmsixpc\ as a function of the
path-length for our two chosen distributions, with the value of the EM
being the one we found for SN~1986J at $t = 20$~yr.

As can be seen, an ionized region near the reverse shock cannot
produce our value of EM except in the case of large mass
\Mion\ concentrated in a very thin shell, for example \Mion =
5~\Msol\ in a shell of thickness only 0.002 times the reverse shock
radius, or $9\times10^{14}$~cm.  Such a large mass within such a very
thin shell seems improbable.

If the ejecta are ionized from the center of the supernova, then a
spherical region of ionized ejecta can easily produce the required
value of EM, with, for example, \Mion = 2~\Msol\ in a region with a
radius of $0.3\times$ that of the reverse shock, or $1.3 \times 10^{17}$~cm.

Our value of EM depends on an assumed electron temperature of $T_e =
10^4$~K. Could the required absorption be produced if $T_e$ were in
fact substantially different?  Lower values of $T_e$ would increase
the absorption, and thus lower the ionized mass, i.e.\ the value of
\Mion, required, however, such low values would likely not be
associated with sufficient ionization to produce the free electrons
required for the absorption.  Higher values of $T_e$ would only
increase the value \Mion\ required.

The value of the EM from the ejecta is very sensitive to the radial
distribution of the ionized ejecta.  The density distribution in the
interior part of the ejecta is much flatter than the steep profile in
the outside regions, but nonetheless probably not constant with
radius.  Density distributions with $\rho \propto r^n$ with $n \sim
-1$ are expected \citep[e.g.,][]{MatznerM1999a}.  For any density
distribution with $n < -0.5$, the formal value of the EM becomes
infinite near $r = 0$.  For inner radii small compared to the outer
radius, however, our conclusions above are not much altered: if, for
example, we take a $\rho \propto r^{-1}$ density distribution
extending inward from $0.25\times$ the reverse shock radius ($1.9
\times 10^{17}$~cm at $t = 20$~yr) to the radius we measured for the
central component of $6.7 \times 10^{16}$~cm, we find that our fitted
value of EM can be produced with $\Mion = 1.5$~\Msol.

Our conclusion, that producing the value of the EM we observed for the
central component requires that the absorbing ionized material be
concentrated in the inner part of SN~1986J, therefore seems robust.
Unless \Mion\ is so concentrated, unreasonably large total masses
($>5$~\Msol) are required to produce the observed absorption.  The
only other way to produce the observed absorption would be to have a
few \Msol\ distributed in a very thin shell near the reverse shock
which we consider unlikely, or of course to have \Mion\ not
spherically distributed, which would imply that our particular line of
sight suffers from much higher absorption than most.  Our value of EM
also places some constraints on the associated dispersion measure, but
we defer discussion of dispersion measure and the implications for
fast radio bursts to a companion paper: \citet{SN86J-FRB}.

\subsection{Is the Central Component Due to a Common-Envelope Evolution of SN~1986J's Progenitor?}
\label{sbinarySN}

The first hypothesis for origin of the bright central component was
recently suggested by \citet{Chevalier2012b}.  In this hypothesis, Type
IIn supernovae, such as SN~1986J, arise from massive-star binary
systems that have undergone a period of common-envelope (CE) evolution
after the first star in the system exploded as a supernova.  SNe
resulting from massive binary systems are not unexpected since a large
fraction of massive stars is expected to be in binaries
\citep{Sana+2012}.  The explosion of the first star, not observed in
the case of SN~1986J, leaves behind a compact object (neutron star or
black hole), and a subsequent CE phase results in the inspiral of the
compact object into the second star.  The CE phase causes a period of
very strong and highly anisotropic mass loss, which is followed by the
second, observed supernova explosion \citep[for a more elaborate,
  triple-star version of this scenario, see][]{JusthamPV2014}.  The
result is a highly anisotropic CSM, much denser in the equatorial
plane of the binary than elsewhere.

If SN~1986J originates from such a system, then after the explosion,
part of the shock expands relatively rapidly through the lower density
polar regions, producing the observed ``shell'' component, but another
part of the shock interacts with the very dense equatorial CSM,
expanding much more slowly and producing the observed central
component.  The density contrast between the equatorial and the polar
CSM can be more than an order of magnitude \citep{RickerT2012}.  The
shock has something like an hourglass shape with the central component
representing the narrow waist of the hourglass, with our line of sight
being intermediate between a polar and an equatorial one.  The ejecta
are initially opaque to radio waves, so the ejecta which expand above
and below the equatorial disk will initially free-free absorb any
emission from the part of the shock in the disk.  Only after some time
do the bulk of the ejecta expand, fragment or recombine to the point
where the central disk can be seen, at which point the radio central
component ``turns on.''

Is this hypothesis consistent with our observations?  We showed in
\S~\ref{sabs} above that the observed amount of absorption of the
central component, i.e., the part of the shock interacting with the
putative dense equatorial disk, was compatible with what would be
expected from a few \Msol\ of ionized supernova ejecta.  Although the
shock and the ejecta have a non-spherical structure, the system might
be expected to expand in an approximately homologous manner, albeit
with the portion of the shock in the dense equatorial disk traveling
much more slowly than the rest.  Our observations of the evolving SED
are in fact approximately consistent with a homologously expanding
system. We also find that the emission from the central component and
from the shell have approximately the same optically thin spectral
index, which is also expected in this case.

We further found in \S~\ref{sabs} that the ionized material
responsible for the absorption seen towards the central component is
most likely near the center of the SN, rather than being out towards
the outer edge of the SN.\@ Such concentration near the center is also
expected in the case of a highly anisotropic CSM resulting from CE
evolution, since the dense, ionized material would also be near the
narrow waist of the hourglass.

The CE hypothesis therefore seems broadly consistent with our
observations, although modeling of the CE scenario, against which we
could compare our observations in detail, has not been carried out to
date, and none of our VLBI images obviously suggest an hour-glass
morphology.

\pagebreak[4]
\subsection{Is the Central Component a Pulsar Wind Nebula?}
\label{spwn}

The second hypothesis is that the central component is an emerging
pulsar wind nebula (PWN), forming around the neutron star left behind
in the supernova explosion.  In this case, the central location
and its low proper motion are readily explained.

A PWN is expected to expand with velocities in the range of 1000 to
2000 \kms, and to have a size of 5 to 20\% of that of the forward
shock at ages of a few decades.  This size is consistent with our
measurement of the central component's radius (HWHM) of $\sim$9\% of
the shell radius at $t = 31.6$~yr \citepalias{SN86J-3}.

The 20-GHz flux density of the central component at $t=30$~yr was
$\sim$2 mJy (see Figure \ref{fsed}), corresponding to a spectral
luminosity of $\sim3.3 \times10^{26}$~erg~s$^{-1}$~Hz$^{-1}$, or a
luminosity ($\nu L_\nu$) of $6.6 \times 10^{36}$~erg~s$^{-1}$, which
is $\sim$120 times the current value for the $\sim$1000-yr old Crab
Nebula \citep[see, e.g.,][]{crab-1997}.  The spindown power of young
pulsars is expected to be $\gtrsim 10^{39}$~erg~s$^{-1}$, several
orders of magnitude higher than the above luminosity, and thus easily
able to power the putative PWN.  The spindown power is expected to be
constant or decay only slowly with time \citep[e.g.,][]{KoteraPO2013,
  ChevalierF1992}.  \citet{GelfandSZ2009} modeled the evolution of a
PWN and predicted radio luminosities comparable to that observed for
SN~1986J's central component at $t=20$~yr.  In their model, the radio
luminosity of the PWN decays relatively slowly, approximately as
$t^{-0.5}$, for the first few decades.  Although we indeed found that
the radio luminosity of the central component is decaying with time,
it seems to be decaying significantly faster ($S_\scomp \propto
t^{-2.07\pm0.16}$) than suggested by \citet{GelfandSZ2009}'s model of
PWN.

The optically-thin spectral index of the central component was $-0.76
\pm 0.07$, which is considerably steeper than observed in any known
PWN, which have spectral indices in the range $-0.3$ to 0.0
(\citealt{GaenslerS2006}; see also \citealt{Green2014}).
We further found that the decrease of the absorption of the central
component suggested a system that was decelerated perhaps slightly
more than the shell, which is already significantly decelerated in
SN~1986J \citepalias[with $r \propto t^{0.69}$][]{SN86J-2}.  One would
expect a PWN, by contrast, to be somewhat {\em accelerated}, since the
PWN is driven into the still freely expanding ejecta interior to the
decelerated forward shock \citep[e.g.,][]{ChevalierF1992,
  GelfandSZ2009, Crab_expand2015}.

On the balance, the observations do not favor a PWN explanation,
although little enough is known about young PWN that such an
explanation cannot be decisively ruled out.

\subsection{Is the Central Component Due to a Newly Formed Black Hole?}
\label{sblackhole}

The third hypothesis is that the central component is due to a newly
formed black hole.  The progenitor of SN~1986J was thought to be
massive \citep{Rupen+1987, WeilerPS1990} and might therefore have
produced a black hole instead of a neutron star.  Could the central
component be radio emission originating from a black hole environment,
specifically in the form of jets powered by fallback accretion?  We
note that the accretion could occur onto a neutron star as well as
onto a black hole but, as we explain below, we consider this
possibility less likely.

Radio emitting jets in accreting black-hole systems can have a wide
range of radio spectra, due to the possible presence of both
self-absorbed and optically thin components, so the central
component's observed optically thin spectrum can easily be
accommodated by this hypothesis.  The slightly higher short-term
variability at 22-GHz noted in \S~\ref{svariability} is also expected
in this case.

However, SN~1986J's central component is far more radio-luminous than
any known stellar-mass black hole system: at $t=30$~yr, the central
component has a measured radio luminosity of 
\LR = $3 \times 10^{35}$~erg~s$^{-1}$ ($\nu L_\nu$), with the
unabsorbed value being about an order of magnitude larger
(\S~\ref{sabs}).  Known accreting stellar-mass black hole systems in
X-ray binaries have $\LR \lesssim 10^{32}$~erg~s$^{-1}$ \citep[see,
  e.g.,][]{KordingFC2006}.
Of course, an accreting black hole could have a wide range of $\LR$
depending on the accretion rate, and one in SN~1986J may be accreting
at a much higher rate than any of those in X-ray binaries.

The X-ray emission in accreting black holes is generally correlated
with the radio emission.  Can we therefore constrain a possible
accreting black hole in SN~1986J using X-ray observations?  Indeed, in
the case of the Type IIL SN~1979C, a flattening of the X-ray
lightcurve at a luminosity comparable to $L_{\rm Edd}$ has been
interpreted in terms of a newly formed black hole
\citep{PatnaudeLJ2011}.  However, no clear central component has been
seen in the radio in SN~1979C \citep{SN79C-shell}.  Furthermore,
\citet{DwarkadasG2012} show that SN~1979C's flat X-ray lightcurve, as
well as those of some Type IIn SNe, could be due to circumstellar
interaction rather than a central black hole.

In any case, the X-ray lightcurve of SN~1986J does not seem to be
particularly flat: X-ray observations have been obtained at various
times between $t \simeq 8.5$~yr and 20.6~yr (1991 and 2003), and
\citet{Houck2005a} found that the X-ray flux (0.5 to 2.5 keV) was
declining with time $\propto t^{-1.7 \pm 0.25}$, consistent with CSM
interaction.\footnote{Note that \citet{TempleRS2005}, using the same
  observations up to $t \simeq 19.4$~yr find a significantly {\em
    steeper}\/ decay, $\propto t^{-2.89 \pm 0.19}$.}  No more recent
X-ray measurements have been published.

\citet{Houck2005a} found that the unabsorbed X-ray luminosity, \LX, at
$t = 20.6$~yr was $1.9\times10^{39}$~erg~s$^{-1}$ (0.5 to 2.5 keV),
equal to the Eddington luminosity $L_{\rm Edd}$ of a 16~\Msol\ black
hole.  While this value is possible for an accreting black hole, it is
larger than expected for long-lasting fallback accretion
\citep{Perna+2014}.

It has been observed that the \LR, \LX, and the black-hole mass for
accreting black holes are related and lie on a ``fundamental plane''
\citep[see, e.g.,][]{Ho2008, FalckeKM2004, MerloniHD2003}.  The
relationship holds for a wide range of masses from stellar mass black
holes in Galactic X-ray binaries through supermassive ones in the
centers of galaxies.  If the emission from the center of SN~1986J is
due to an accreting black hole, can we place it on this fundamental
plane?  \citet{MerloniHD2003} gives a plot of the projected
fundamental plane, with an X-axis of $0.60 \log(\LX) + 0.78 \log(M)$,
where \LX\ is in erg~s$^{-1}$ and $M$ is the black hole mass in \Msol.
If we assume that all the X-ray emission at $t \sim 20.7$~yr is due to
the central component, its \LX\ would be $1.9 \times
10^{39}$~erg~s$^{-1}$.  If we further take a relatively high
black hole mass of 20~\Msol, we find that the fundamental plane
relationship predicts that $\LR \simeq 3\times10^{31}$~erg~s$^{-1}$.
This is almost 4 orders of magnitude fainter than the measured \LR\ of
the central component at that age ($\sim 3 \times
10^{35}$~erg~s$^{-1}$).  In fact, given that there is significant
absorption by the ejecta at 5~GHz, the intrinsic \LR\ is likely higher
by up to an order of magnitude.  A mass much above 20~\Msol\ also
seems unlikely.  Although there is considerable scatter in the
fundamental plane relationship, SN~1986J's central component seems to
have a far higher ratio of \LR /\LX\ than accreting black holes which
are on the fundamental plane, in fact far higher than any known
stellar-mass black hole systems.
It might be expected that a black hole as young as one in SN~1986J
would accrete at a much higher rate than the much older Galactic X-ray
binaries, but high accretion rates seem to {\em quench}\/ the radio
emission rather than enhance it \citep[e.g.,][]{FenderGR2010}.
SN~1986J's high radio luminosity, and in particular, its high ratio of
\LR /\LX, seem discrepant with that seen in other accreting black hole
systems.

As noted above, the accreting object could in fact be a neutron star
rather than a black hole, as accreting neutron stars can also power
jets and produce radio emission. However, for the same X-ray
luminosity, neutron-star systems tend to have lower radio luminosities
than black-hole systems \citep{Kording2014}, so we consider an
accreting neutron star less likely than a black hole.

In summary, the clearly distinct, stationary central compact radio
component could be interpreted in terms of an accreting black-hole
system.  As we noted in \citetalias{SN86J-3}, the hotspot to the NE is
approximately aligned with the extension of the central component,
which, although quite inconclusive, could be interpreted in terms of a
jet structure.  However, although they do not rule out the black-hole
hypothesis, the high radio luminosity, the declining X-ray lightcurve
and the large ratio of $\LR/\LX$ argue against it.

\section{Summary and Conclusions}

We obtained new multi-frequency VLA flux density measurements of
SN~1986J, made in conjunction with VLBI observations, which showed the
continued evolution of its spectral energy distribution.  We found
that:

\begin{trivlist}

\item{1.} The radio spectral energy distribution shows both an
  inversion and a high frequency turnover.  Between these two points,
  the emission increases with increasing frequency.  The inverted part
  of the spectrum is due to the central component, which is absorbed
  at low frequencies thus producing the high-frequency turnover.  At
  frequencies below the inversion, the shell emission dominates.  The
  inversion is presently near 2~GHz, and the high-frequency turnover
  near 15~GHz, and both the inversion and turnover are progressing
  downward in frequency with time.

\item{2.} We performed a Bayesian fit of a model consisting of a shell
  and a partly absorbed central component to the evolving spectrum.
  The fit suggests that the optically thin, or intrinsic, spectral
  indices of the central component and the shell radio emission are
  consistent with being the same (at $\alpha \simeq -0.64$).  At $t =
  20$~yr, the absorption of the central component is produced by
  material with an emission measure of $(1.64 \pm 0.20) \times
  10^9$~\cmsixpc\ which is decreasing with time $\propto t^{-2.72 \pm
    0.26}$.  The unabsorbed flux densities of both the central
  component and the shell are decreasing with time, although that of
  the central component is decreasing less rapidly, hence the central
  component becomes ever more dominant.  The evolution of the spectrum
  is approximately consistent with a system with a constant mass and
  ionization fraction expanding homologously with the same $r \propto
  t^{0.69}$ that is observed for the shell with VLBI.

\item{3.} The central component is unlikely to be due to a dense
  condensation in the CSM, central only in projection, due to its high
  surface brightness and its long lifetime.  It is therefore almost
  certainly in the physical center of the expanding shell of ejecta.
\item{4.} The ionized material responsible for the radio absorption
  seen towards the central component must be distributed
  non-uniformly.  To provide the necessary absorption with less then
  5~\Msol\ of ionized material in a spherically-symmetric distribution
  probably requires that the ionized material be concentrated towards
  the center of SN~1986J, and not near the reverse shock.

\item{5.} The central component could be explained by SN~1986J being
  the second supernova in a binary system, with SN~1986J's ejecta
  interacting with an aspherical region of very dense mass-loss from
  the progenitor which arose during period of common-envelope binary
  evolution when the compact remnant of the first supernova was within
  the envelope of SN~1986J's progenitor.
  The central component in this picture is due to the part of
  the supernova shock which is interacting with the disk of very dense
  CSM produced resulting from the common-envelope phase, while the
  remainder of the shell is due to less equatorial parts of the shock
  which interact with the more normal supergiant wind.

\item{6.} The central component could be due to a young pulsar wind
  nebula, however, the steep intrinsic spectrum and the relatively
  rapid decay with time of its unabsorbed flux density is not
  consistent with what is expected.

\item{7.} The central component might be due to a newly-formed
  accreting black hole in the center of SN~1986J\@.  However, it has a
  far higher radio luminosity, both absolutely and in comparison to
  the X-ray luminosity, than any known stellar-mass black-hole
  systems.  On the other hand, not much is known
  about the characteristics of newly-formed black holes in supernovae.

\end{trivlist}

\section*{Acknowledgments }

We have made use of NASA's Astrophysics Data System Bibliographic
Services, as well as the NASA/IPAC Extragalactic Database (NED) which
is operated by the Jet Propulsion Laboratory, California Institute of
Technology, under contract with the National Aeronautics and Space
Administration.  This research was supported by both the National
Sciences and Engineering Research Council of Canada and the National
Research Foundation of South Africa.

\software{AIPS \citep{Greisen2003},
  CASA \citep{McMullin+2007},
  PyMC \citep[v2.3,][]{Fonnesbeck+2015},
  corner.py \citep{Foreman-Mackey2016}}

\appendix
\section{Details of Bayesian Fit and Comparison to Least-Squares Fit}
\label{aBaylsq}

We performed a Bayesian analysis and fitted a simple model of the evolution of
the SED to our flux-density measurements.  The model has eight free
parameters, described in \S~\ref{smodel}, namely $S_{0,\sshell}, \,
b_\sshell, \alpha_\sshell, \, S_{0,\,\scomp}, \, b_\scomp, \,
\alpha_\scomp$, EM$_0$, and $b_\sEM$.  We performed a Monte-Carlo
Markov chain integration to estimate the posterior probability
distribution of the free parameters given our flux density
measurements (both ones listed in Table \ref{tvla} and earlier ones).
We used the PyMC package, version 2.3, by C. Fonnesbeck et al., which
is available at \url{http://pymc-devs.github.io/pymc/README.html}.  We
give the prior distributions used for each parameter in
Table~\ref{tBaylsq}.

\begin{figure*}[h]
\centering
\includegraphics[width=\linewidth]{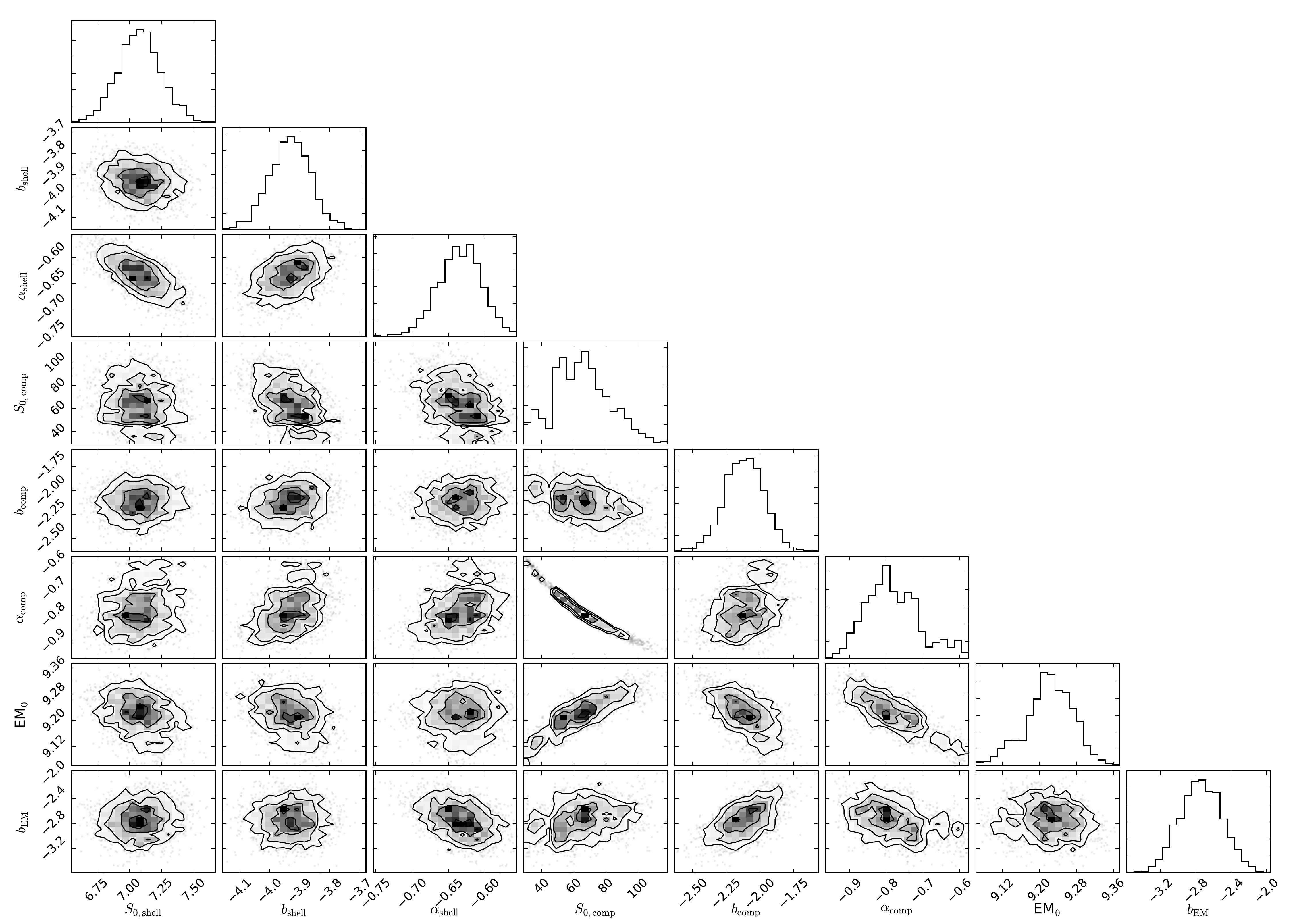}
\caption{Joint posterior distribution (``corner plot'') of the fitted
  parameters.  The top plot in each column gives the marginal
  posterior distribution of the fitted variable, while the remainder
  show scattergrams showing the correlation between pairs of the
  fitted parameters.  The contours in the scatter plots are drawn at
  the 1, 2, and 3$\sigma$ points in the distribution. The plot was
  made with the ``corner.py'' package \citep{Foreman-Mackey2016}.}
\label{fcorner}
\end{figure*}

The mean values and standard deviations over the posterior
distributions were given in \S~\ref{sBresults}, but we repeat them in
Table~\ref{tBaylsq} for convenience. We show the details of the
posterior distributions in a ``corner plot'' in Figure~\ref{fcorner}.
We also performed a weighted least-squares fit of the model to the
measurements, and we obtained the results also listed in
Table~\ref{tBaylsq}.  The means over the Bayesian posterior
distributions, which we use as our estimators, are consistent well
within $1\sigma$ with the least-squares best-fit values, as is
expected since our prior distributions are largely uninformative.

The fitted values for $S_{0,\scomp}$ and $\alpha_\scomp$ are strongly
anti-correlated, but this is largely due to our use of a low nominal
frequency of 1~GHz for $S_{0,\scomp}$.  Since the low-frequency
flux-density is dominated by that of the shell, our measurements only
poorly constrain the flux density of the central component at 1~GHz
($S_{0,\scomp}$).  At 10~GHz, where the central component is more
dominant, the measurements constrain the flux density of the central
component much more accurately.  Therefore low values of
$S_{0,\scomp}$ in conjunction with flat values of $\alpha_\scomp$, or
high values of $S_{0,\scomp}$ in conjunction with steep values of
$\alpha_\scomp$ are both possible.  The flux density of the central
component at 10~GHz is thus more narrowly constrained than our value
of $S_{0,\scomp}$ at 1~GHz of $61\pm 17$~mJy might suggest: examining
the posterior distribution of the central component's flux density at
10~GHz, we obtain a mean value of $10.4 \pm 1.1$~mJy.

\begin{deluxetable*}{L l L L l}[h]
\tablecaption{Comparison of Bayesian and Least-Squares Estimates\label{tBaylsq}}
\tablehead{
  \colhead{Parameter} & \multicolumn{1}{c}{Prior Distribution} &
  \multicolumn{1}{l}{Bayesian fit posterior mean} &
    \colhead{Least squares best-fit} & \multicolumn{1}{l}{Units} \\[-7pt]
  &  & \multicolumn{1}{c}{and standard deviation} 
     & \multicolumn{1}{c}{and standard error} }
\decimals
\startdata
S_{0,\sshell}   & uniform, 3 to 20      & \phneg 7.07 \pm 0.17 & \phneg 7.08 \pm 0.17 & mJy \\
b_\sshell      & uniform, $-5$ to $-2$ &       -3.92 \pm 0.07 & -3.92 \pm 0.08 & \nodata \\
\alpha_\sshell & normal, mean=$-0.6, \sigma=1$
                                      &  -0.63 \pm 0.03 & -0.63 \pm 0.03 & \nodata \\
S_{0,\scomp}    & uniform, 5 to 200     & \phneg  61 \pm 17     & \phneg 60 \pm 19   & mJy \\
b_\scomp       & uniform, $-4$ to +1   &  -2.07 \pm 0.16 & -2.11 \pm 0.16 & \nodata \\
\alpha_\scomp  & normal, mean=$-0.6, \sigma=1$
                                      &  -0.76 \pm 0.07 & -0.77 \pm 0.08 & \nodata \\
{\rm EM}_0    & log-uniform, 0.1 to 10 &\phneg 1.64 \pm 0.20 & \phneg 1.64 \pm 0.22 & $\times 10^9$ \cmsixpc \\
b_\sEM         & uniform, $-5$ to $-1$ &  -2.72 \pm 0.26& -2.76 \pm 0.26 & \nodata \\
\enddata
\end{deluxetable*}

\clearpage

\bibliographystyle{aasjournal} 

\bibliography{mybib1}

\end{document}